\documentclass{ws-p8-50x6-00}

\begin{document}

\title{Proton stability in 5D GUTs with orbifold compactification}

\author{Archil Kobakhidze}

\address{HEP Division, Department of Phydics, University of Helsinki and \\ 
Helsinki Institute of Physics, FIN-00014 Helsinki, Finland \\ 
Andronikashvili Institute of Physics, GE-380077 Tbilisi, Georgia \\
E-mail: Archil.Kobakhidze@helsinki.fi}


\maketitle

\abstracts{
We construct SU(5) SUSY GUT in 5D compactified on $S^1/Z_2$ orbifold 
where the matter fields are living in the five dimensional bulk. SU(5) symmetry is broken 
down to the Standard Model gauge group by the orbifold projection which 
automatically ensures stability of proton in all orders of perturbation theory. 
The model predicts extra mirror quark-lepton families 
which along with the GUT particles and the excitations
of extra dimensions could be observable at high energy colliders providing
the unification scale is in the TeV range.}

Recently, various issues of four dimensional (4D) particle phenomenology 
have been reconsidered within the higher dimensional theories. The models with 
large extra dimensions are particularly interesting from phenomenological and experimental 
point of view \cite{1,2}. Typically such models suffer from the 
potentially large and thus phenomenologically unacceptable
violation of certain global symmetries. 
A large flavour changing neutral currents, large neutrino masses, unacceptably fast proton
decay etc. could be indueced in the low-energy  effective theory. 
In this note we will concentrate on the proton stability problem
in higher dimensional GUTs with low scale unification \cite{3}.

Let us consider supersymmetric $SU(5)$ model in higher-dimensional
space-time. For simplicity we restrict our discussion by the case of one
compact extra space-like dimension. 
As usually, each family of ordinary quarks and leptons is placed in $%
\overline{5}$ and $10$ irreducible representations of the $SU(5)$ group but
now the ordinary quarks and leptons are supplemented by the mirror states.
The mirror states combine with ordinary quarks and leptons to  form $N=2$
hypermultiplets: ${\cal Q}=\left( Q_{L},Q_{R}\right) \sim \overline{5}$, 
${\cal D}=\left( D_{L},D_{R}\right) \sim 10$. 
Here $Q_{L}(D_{L})$ and $Q_{R}(D_{R})$ are $N=1$ left-handed and
right-handed chiral quintuplet (decuplet) superfields, respectively. 
$N=2$ $SU(5)$ vector supermultiplet ${\cal V}=\left( V,\Phi \right)$
contains $N=1$ 4-dimensional gauge superfield $V=\left( A^{\mu },\lambda
^{1},X^{3}\right) $ as well as $N=1$ chiral superfield $\Phi =\left( \Sigma
+iA^{5},\lambda ^{2},X^{1}+iX^{2}\right) $ both in the adjoint
representation of the $SU(5)$ group. Finally, we also need to introduce at least one $SU(5)$
fundamental and one anti-fundamental hypermultiplets: ${\cal H}=(H_{L},H_{R})\sim 5$, 
$\widetilde{{\cal H}}=(\widetilde{H}_{L},\widetilde{H}_{R})\sim \overline{5}$, 
where the electroweak Higgs doublet (anti-doublet) $%
h_{L(R)}^{W}$ ($\widetilde{h}_{L(R)}^{W}$) presumably resides. One obvious
advantage of the $N=2$ supersymmetric GUTs is that the gauge fields in $V$
and scalars in $\Phi $ are unified in the same $N=2$ vector supermultiplet 
${\cal V}$. The scalar component of the chiral superfield $\Phi $ can be used
to break $SU(5)$ gauge symmetry. Alternatively, one can break $%
SU(5)$ invariance through the orbifold compactification.
\begin{table}[t] \centering%
\begin{tabular}{lll}
\hline
Fields &  & $Z_{2}$ parity, ${\cal P}$ \\ \hline
Vector supermultiplet, ${\cal V}$ & $\left\{
\begin{array}{l}
G(\Phi _{G})\sim \left( 8,1,0\right) \\
W(\Phi _{W})\sim \left( 1,3,0\right) \\
S(\Phi _{S})\sim \left( 1,1,0\right) \\
X,(\Phi _{X})\sim \left( \overline{3},2,-\sqrt{\frac{5}{12}}\right) \\
\overline{X}(\Phi _{\overline{X}})\sim \left( 3,2,\sqrt{\frac{5}{12}}\right)
\end{array}
\right. $ & \multicolumn{1}{c}{$
\begin{array}{c}
+(-) \\
+(-) \\
+(-) \\
-(+) \\
-(+)
\end{array}
$} \\ \hline
Quintuplets, ${\cal Q}$, $\widetilde{{\cal Q}}$ & $\left\{
\begin{array}{c}
\overline{d}_{L(R)},\widetilde{\overline{d}}_{L(R)}\sim \left( \overline{3}%
,1,-\sqrt{\frac{2}{15}}\right) \\
l_{L(R)},\widetilde{l}_{L(R)}\sim \left( 1,\overline{2},\sqrt{\frac{3}{20}}%
\right)
\end{array}
\right. $ & \multicolumn{1}{c}{$
\begin{array}{l}
-(+),+(-) \\
+(-),-(+)
\end{array}
$} \\ \hline
Decuplets, ${\cal D}$, $\widetilde{{\cal D}}$ & $\left\{
\begin{array}{c}
\overline{u}_{L(R)},\widetilde{\overline{u}}_{L(R)}\sim \left( \overline{3}%
,1,\frac{2}{\sqrt{15}}\right) \\
q_{L(R)},\widetilde{q}_{L(R)}\sim \left( 3,2,-\frac{1}{\sqrt{60}}\right) \\
\overline{e}_{L(R)},\widetilde{\overline{e}}_{L(R)}\sim \left( 1,1,-\frac{3}{%
\sqrt{15}}\right)
\end{array}
\right. $ & \multicolumn{1}{c}{$
\begin{array}{c}
-(+),+(-) \\
+(-),-(+) \\
-(+),+(-)
\end{array}
$} \\ \hline
Higgs hypermultiplets, ${\cal H},\widetilde{{\cal H}}$ & $\left\{
\begin{array}{c}
h_{L(R)}^{C}\sim \left( 3,1,\sqrt{\frac{2}{15}}\right) \\
h_{L(R)}^{W}\sim \left( 1,2,-\sqrt{\frac{3}{20}}\right) \\
\widetilde{h}_{L(R)}^{W}\sim \left( 1,2,\sqrt{\frac{3}{20}}\right) \\
\widetilde{h}_{L(R)}^{C}\sim \left( \overline{3},1,-\sqrt{\frac{2}{15}}%
\right)
\end{array}
\right. $ & \multicolumn{1}{c}{$
\begin{array}{l}
-(+) \\
+(-) \\
+(-) \\
-(+)
\end{array}
$} \\ \hline
\end{tabular}
\caption{$Z_2$-parities ${\cal P}$  of various fields.}%
\end{table}%
We consider here the later possibility by  
compactifying the extra fifth dimension on an $S^{1}/Z_{2}$ orbifold \cite{2,4}. What 
we will require additionally is the conservation of $B$ and/or $L$ global
charges upon the compactification. In fact this can be  achieved rather naturally.
First, in addition to the particle content given above, we have to introduce an 
extra quintuplet $\widetilde{{\cal Q}}$ $\sim \overline{5}$and an
extra decuplet $\widetilde{{\cal D}}$, $\sim 10$ of matter fields per each
family of quarks and leptons. The second step is to appropriately project
the different states in ${\cal Q}$, $\widetilde{{\cal Q}}$, ${\cal D}$, and $%
\widetilde{{\cal D}}$ upon the dimensional reduction. This can be
done by assigning different orbifold $Z_{2}$-numbers to
the quarks and leptons (and
their mirrors) in ${\cal Q}$, $\widetilde{{\cal Q}}$, ${\cal D}$, and $%
\widetilde{{\cal D}}$ as it is given in Table 1. According to the Fourier expansion of 5D fields, 
the wave functions of the parity-odd fields vanish at the
orbifold fixed-points ($y=0,\pi $) and only parity-even fields can propagate
on the 4-dimensional boundary walls. This suggest the identification of
ordinary quarks and leptons with ${\cal Q}$,$\widetilde{{\cal Q}}$, ${\cal D}
$ $\widetilde{{\cal D}}$ fragments as: $\widetilde{\overline{d}}_{L},\widetilde{\overline{u}}_{L},q_{L},l_{L},%
\widetilde{\overline{e}}_{L}$. There also present their mirror states on the boundary wall:
$\overline{d}_{R},\overline{u}_{R},\widetilde{q}_{R},\widetilde{l}_{R},%
\overline{e}_{R}$. The gauge symmetry on the fixed point $y=0$ is just $SU(3)_{C}\otimes SU(2)_{W}\otimes
U(1)_{Y}$ one and $N=2$ supersymmetry is reduced to the $N=1$. Beside the mirror states we
have some additional states beyond the usual particle content of the MSSM 
as one can see from Table 1.

Now, since the gauge superfields $X,\overline{X}$ are $Z_{2}$-odd, they
decoupled from the zero modes of quarks and leptons and thus they can not be
responsible for the $B$ and $L$ violating interactions among them anymore.
The adjoint scalar superfields $\Phi _{X}$ and $\Phi _{\overline{X}}$,
contrary, have a zero modes and couple to the matter on the boundary wall.
However, they can only transform the mirror quarks into the ordinary leptons
and the mirror leptons into the ordinary quarks. 
So among the possible final states along with the ordinary quarks and
leptons always will appear the mirror ones. This actually means that in the
limit of massless ordinary quarks and leptons and their mirror partners the
following global charges are separately conserved: $Q_{1}=B+L_{M},  Q_{2}=L+B_{M}$, 
where the mirror baryon and lepton numbers we denote as $B_{M}$ and $L_{M}$,
respectively. To generate the masses for the ordinary quarks and leptons and
their mirror partners we add $N=2$ supersymmetry violating Yukawa terms 
on the boundary wall at $y=0$ \cite{3}. These Yukawa terms contain only zero modes 
of the fields and thus the colored Higgs triplets from $%
H_{L}$ and $\widetilde{H}_{L}$ completely decoupled from the quarks and
leptons as well as from their mirror partners. This is intrinsically geometric 
mechanism for the doublet-triplet splitting. Thus, the
Yukawa interactions on the boundary wall respect the global charges 
$Q_{1}$ and $Q_{2}$. It is evident
now that, since the mirror particles is assumed to be heavier
than the ordinary ones, the proton decay is forbidden
kinematically. In other words, as long as mirror particles cannot
be produced $B$ and $L$ are separately conserved. As a result the
proton is absolutely stable in all orders of perturbation theory.

\end{document}